\documentclass[aps,prd,preprint,superscriptaddress,showpacs]{revtex4}

\def\beq{\begin{equation}}
\def\eeq{\end{equation}}
\def\bea{\begin{eqnarray}}
\def\eea{\end{eqnarray}}
\def\nn{\nonumber}
\def\roughly#1{\mathrel{\raise.3ex\hbox
{$#1$\kern-.75em\lower1ex\hbox{$\sim$}}}}

\def\sss{\scriptscriptstyle}

\def\bd{B_d^0}
\def\bdbar{{\overline{B_d^0}}}

\def\ks{K_{\sss S}}

\begin{document}

\preprint{UdeM-GPP-TH-05-139}

\title{\boldmath Measurements of New Physics in $B \to \pi \pi$ Decays}

\author{Seungwon Baek}
\affiliation{Physique des Particules,
        Universit\'{e} de Montr\'{e}al,
	C.P.\ 6128, succ.\ centre-ville,
	Montr\'{e}al, QC, Canada H3C 3J7}
\author{F.\ J.\ Botella}
\affiliation{Departament de F\'{\i}sica Te\`{o}rica and IFIC,
        Universitat de Val\`{e}ncia-CSIC,
        E-46100, Burjassot, Spain}
\author{David London}
\affiliation{Physique des Particules,
        Universit\'{e} de Montr\'{e}al,
	C.P.\ 6128, succ.\ centre-ville,
	Montr\'{e}al, QC, Canada H3C 3J7}
\author{Jo\~{a}o P.\ Silva}
\affiliation{Centro de F\'{\i}sica Te\'{o}rica de Part\'{\i}culas,
	Instituto Superior T\'{e}cnico,
	P-1049-001 Lisboa, Portugal}
\affiliation{Instituto Superior de Engenharia de Lisboa,
	Rua Conselheiro Em\'{\i}dio Navarro,
	1900 Lisboa, Portugal}

\date{\today}

\begin{abstract} If new physics (NP) is present in $B\to\pi\pi$ decays,
it can affect the isospin $I=2$ or $I=0$ channels. In this paper, we
discuss various methods for detecting and measuring this NP. The
techniques have increasing amounts of theoretical hadronic input. If
NP is eventually detected in $B\to\pi\pi$ --- there is no evidence for
it at present --- one will be able to distinguish $I=2$ and $I=0$, and
measure its parameters, using these methods.
\end{abstract}


\maketitle

\section{Introduction}

The decay $\bd\to\pi^+\pi^-$ was originally thought to be dominated by
a single weak decay amplitude, the tree amplitude. As such, it could
be used to extract the Cabibbo-Kobayashi-Maskawa (CKM) angle $\alpha$
\cite{pdg}. However, it was shown that this decay also receives
significant contributions from the penguin amplitude, which has a
different weak phase \cite{penguins}. This is referred to as ``penguin
pollution.'' Because of this, $\alpha$ cannot be obtained cleanly from
measurements of $\bd(t) \to\pi^+\pi^-$.

However, in 1990 it was shown that it is possible to use isospin to
remove the penguin pollution and extract $\alpha$ \cite{isospin}.
Briefly, the argument goes as follows. In $B\to\pi\pi$ decays, the
final $\pi\pi$ state can have only isospin $I=0$ or $I=2$. Thus,
there are only two decay paths, implying that the amplitudes for the
three decays $B^+ \to \pi^+ \pi^0$, $\bd\to\pi^+\pi^-$ and
$\bd\to\pi^0\pi^0$ obey a triangle relation. In terms of the isospin
amplitudes $A_0$ ($I=0$) and $A_2$ ($I=2$), the amplitudes can be
decomposed as \cite{isospin}
\bea
\label{isospindecomp}
-\sqrt{2} A(B^+ \to \pi^+ \pi^0) & = & 3 A_2 ~, \nn\\
- A(\bd \to \pi^+ \pi^-) & = & A_2 + A_0 ~, \nn\\
-\sqrt{2} A(\bd \to \pi^0 \pi^0) & = & 2 A_2 - A_0 ~,
\eea
where $A_0$ and $A_2$ include both weak and strong phases. The
amplitudes for the CP-conjugate processes are obtained from the above
by changing the sign of the weak phases. From these expressions, we
see that the triangle relation is $\sqrt{2} A(B^+ \to \pi^+ \pi^0) =
A(\bd \to \pi^+ \pi^-) + \sqrt{2} A(\bd \to \pi^0 \pi^0)$, and
similarly for the CP-conjugate decays.

In the Standard Model (SM), the electroweak-penguin (EWP)
contributions to $B \to \pi\pi$ decays are expected to be very small,
so that the amplitude $A_2$ has a well-defined weak phase: $A_2 =
|A_2| \exp(i \gamma) \exp(i \delta_2)$, where $\gamma$ and $\delta_2$
are the weak and strong phases, respectively. In the CP-conjugate
process, the $I=2$ contribution is thus simply $|A_2| \exp(-i \gamma)
\exp(i \delta_2)$. On the other hand, due to the presence of the
penguin amplitude, $A_0$ does not have a well-defined weak phase. It
can be written $A_0 = |A_0| \exp(i \theta)$, where $|A_0|$ and
$\theta$ are complicated functions of both weak and strong phases. The
$I=0$ contribution to the CP-conjugate processes is therefore $|{\bar
A}_0| \exp(i {\bar\theta})$, where $|{\bar A}_0|$ and ${\bar\theta}$
differ from $|A_0|$ and $\theta$. We can now count the total number of
theoretical parameters. Setting $\delta_2 = 0$ (an overall phase is
not physical), we find there are six: $|A_2|$, $|A_0|$, $|{\bar
A}_0|$, $\theta$, ${\bar\theta}$ and $\gamma$. However, there are also
six independent experimental measurements \cite{footnote1}: three
average branching ratios ($B_{+0}$, $B_{+-}$ and $B_{00}$), two direct
CP asymmetries ($C_{+-}$ and $C_{00}$; the direct CP asymmetry
$C_{+0}$ is expected to vanish in the SM), and one interference CP
asymmetry ($S_{+-}$). (The sub-indices refer to the charges of the
physical pions in the final state.) We can therefore obtain all
theoretical parameters, up to discrete ambiguities. In particular,
combining the phase of $\bd$--$\bdbar$ mixing ($\beta$), we can
extract $\alpha = \pi - \beta - \gamma$.

Obviously, it is also possible to extract the other five theoretical
parameters. This was shown explicitly by Charles \cite{charles}. In
his analysis, a diagrammatic decomposition of the $B\to\pi\pi$
amplitudes was used:
\bea
\label{amps}
-\sqrt{2} A(B^+ \to \pi^+ \pi^0) & = & T + C ~, \nn\\
- A(\bd \to \pi^+ \pi^-) & = & T + P ~, \nn\\
-\sqrt{2} A(\bd \to \pi^0 \pi^0) & = & C - P ~.
\eea
Here, $T$ and $C$ are (nominally) the color-allowed and
color-suppressed tree amplitudes, respectively, while $P$ is the
penguin contribution. All three contain both weak and strong phases.
The one complication here is that, while $T$ and $C$ are both
proportional to $V_{ub}^* V_{ud} \sim \exp(i \gamma)$, the penguin
amplitude contains three contributions, corresponding to the flavor of
the internal quark. These can be written
\beq
P_u V_{ub}^* V_{ud} + P_c V_{cb}^* V_{cd} + P_t V_{tb}^* V_{td} =
(P_u - P_c) V_{ub}^* V_{ud} + (P_t - P_c) V_{tb}^*
V_{td} ~,
\label{Pdefs}
\eeq
where the $P_i$ ($i=u$, $c$, $t$) contain only strong phases. Above,
the unitarity of the CKM matrix has been used to eliminate the $P_c$
term. The $(P_u - P_c)$ term can be absorbed in the definitions of $T$
and $C$, so that
\beq
T = [T_{tr} + (P_u - P_c)] V_{ub}^* V_{ud} ~~,~~~~ C = [C_{tr} - (P_u -
P_c)] V_{ub}^* V_{ud} ~,
\label{mixeddiags}
\eeq
where $T_{tr}$ and $C_{tr}$ denote the pure tree-level color-allowed
and color-suppressed tree contributions, respectively. These contain
only strong phases; the weak phases have been factored out. Thus, the
diagrams $T$ and $C$ of Eq.~(\ref{amps}) actually contain both tree
and $(P_u - P_c)$ penguin contributions, while the penguin diagram $P$
is actually $(P_t - P_c) V_{tb}^* V_{td}$.

With these (re)definitions, all three diagrams of Eq.~(\ref{amps})
have well-defined weak phases -- the phase of $T$ and $C$ is $\gamma$,
while that of $P$ is $-\beta$. Thus, in this parametrization, taking
into account the phase of $\bd$--$\bdbar$ mixing ($\beta$), there are
again six parameters: the magnitudes $|T|$, $|C|$ and $|P|$, two
relative strong phases, and the weak phase $\alpha = \pi - \beta -
\gamma$. As above, it is possible to extract all theoretical
parameters.

Having shown that it is possible to extract $\alpha$ and the remaining
theoretical parameters from $B\to\pi\pi$ decays within the SM using an
isospin analysis, it is natural to ask how things are affected should
physics beyond the SM be present. In particular, we want to ascertain
whether it is possible to detect the new physics (NP) and measure its
parameters. If so, we want to perform the relevant fits using present
data.

We have noted above that there are two isospin amplitudes in
$B\to\pi\pi$, $I=0$ and $I=2$. Thus, if NP is present, it can
contribute to either of these isospin channels. However, in a recent
paper \cite{BBLS}, we showed that if no assumptions are made
concerning hadronic parameters, then the $I=0$ NP amplitude can never
be detected, and only one piece of the $I=2$ NP amplitude can be
seen. The main aim of this paper is to investigate what happens if one
adds hadronic input. In particular, we consider the addition of
theoretical assumptions about the size of $|C/T|$ and $|P/T|$, as well
as the strong phases of $T$ and the NP contributions. (The former
assumptions are on stronger footing than the latter.)

Note that the decompositions in terms of isospin amplitudes ($A_0$,
$A_2$) and diagrams ($T$, $C$, $P$) are equivalent. Throughout this
paper we will use a mixed notation, in which the SM is described by
diagrams, but new physics is described in terms of isospin amplitudes.

In Section 2, we examine the SM explanation of current $B\to\pi\pi$
data, with no theoretical hadronic input. We find a good fit in this
case. In Sec.~3, we add theoretical hadronic estimates of $|C/T|$ and
$|P/T|$. In this case, there is a discrepancy in the SM fit, but only
at the level of about $1.5\sigma$. We then consider the addition of
new physics. If no theoretical input is added (Sec.~4), we again
obtain a good fit. We consider the addition of such input in
Sec.~5. We parametrize the NP using reparametrization invariance. Here
we fit for $I=2$ and $I=0$ NP separately. We find that the addition of
$I=2$ NP gives a fit which is not great, but still acceptable. There
are not enough observables to measure the $I=0$ NP parameters. In
order to make more progress, we change parametrizations in the
following sections. We assume that the NP strong phases are all small
in Sec.~6. We find a good fit for $I=2$ NP, but still cannot fit for
$I=0$ NP. Finally, in Sec.~7, we add a third piece of theoretical
hadronic input, that the phases of $T$ and the NP are equal. In this
case, we are able to measure both $I=2$ and $I=0$ NP parameters,
obtaining good fits in both cases. We conclude in Sec.~8.

\section{Standard Model: No Hadronic Input}

We begin with a review of the standard-model fit to current
measurements. In order to easily compare with the case of new physics
in subsequent sections, our analysis is slightly different from what
is usually presented.

As discussed in the introduction, the $B\to\pi\pi$ measurements alone
allow one to extract the CP phase $\alpha$. However, it is possible to
measure the weak phases independently. The phase $\beta$ has already
been measured very precisely in $\bd(t) \to J/\psi\ks$: $\sin 2\beta =
0.726 \pm 0.037$ \cite{sin2beta}. Note that this value is consistent
with SM expectations. If new physics is present in $B\to\pi\pi$
decays, it could affect $\bd$--$\bdbar$ mixing. However, the data show
that, even if NP is present in $\bd$--$\bdbar$ mixing, its effect on
$\beta$ is minimal. The measurement of $\sin 2\beta$ determines
$2\beta$ up to a twofold ambiguity. Here we assume that $2\beta
\simeq 47^\circ$, or $\beta = 23.5^\circ$, in agreement with the SM.

The phase $\gamma$ can in principle be measured without penguin
pollution through CP violation in decays such as $B\to DK$ \cite{BDK}.
Alternatively, $\gamma$ can be obtained from a fit to a variety of
other measurements, some non-CP-violating. The latest analysis gives
$\gamma = {58.2^{+6.7}_{-5.4}}^\circ$ \cite{CKMfitter}. Note that this
error includes uncertainties in theoretical quantities. For the
purposes of the fits, we assume symmetric errors, and take $\gamma =
(58.2 \pm 6.0)^\circ$.

It is therefore possible to obtain $\beta$ and $\gamma$ (or $\alpha =
\pi - \beta - \gamma$) independently. Here and below, we assume that
the SM weak phases are already known, having been obtained from other
decays. The SM analysis therefore consists of a fit to all
$B\to\pi\pi$ data, along with these independent determinations of the
CP phases. The latest $B\to\pi\pi$ measurements are shown in
Table~\ref{T:data}.
%
\begin{table}
\caption{Branching ratios, direct CP asymmetries $C_f$,
and interference CP asymmetries $S_f$ (if applicable) for the three $B
\to \pi\pi$ decay modes. Data comes from
Refs.~\cite{BRs,ACPs,pi+pi-}; averages (shown) are taken from
Ref.~\cite{HFAG}.}
\begin{ruledtabular}
\begin{tabular}{lccc}
& $BR[10^{-6}]$ & $C_f$  & $S_f$ \\
\hline
$B^+ \to \pi^+ \pi^0$ & $5.5 \pm 0.6$ & $0.02 \pm 0.07$ & \\
$B^0 \to \pi^+ \pi^-$ & $4.6 \pm 0.4$ & $-0.37 \pm 0.10$ & $-0.50 \pm
    0.12$ \\
$B^0 \to \pi^0 \pi^0$ & $1.51 \pm 0.28$ & $-0.28 \pm 0.39$ \\
\end{tabular}
\end{ruledtabular}
\label{T:data}
\end{table}

Performed in this way, the fit yields a single solution for the five
unknown theoretical parameters (Table~\ref{SMfit}). We find
$\chi^2_{min}/d.o.f.\ = 0.27/2$. We therefore see that the SM can
acceptably account for the data (leaving aside the somewhat small
value of $\chi^2_{min}/d.o.f.$).

\begin{table}
\caption{Results of a fit within the SM. Inputs are the $B \to \pi\pi$
data (Table~\ref{T:data}) and independent determinations of the CP
phases. All angles are in degrees.}
\begin{ruledtabular}
\begin{tabular}{ccccc}
$|T|$ & $|C|$ & $|P|$ & $\delta_{\sss C} - \delta_{\sss T}$ &
$\delta_{\sss P} - \delta_{\sss T}$ \\
\hline
$21.8 \pm 1.1$ & $18.5 \pm 1.9$ & $5.6 \pm 1.9$ & $-66.6 \pm 13.6$ &
$-52.9 \pm 20.7$ \\
\end{tabular}
\end{ruledtabular}
\label{SMfit} 
\end{table}

\section{Standard Model: Hadronic Input}

There is one potentially worrisome aspect about the fit described in
the previous section: naive theoretical estimates of the relative
sizes of the magnitudes of the amplitudes yield $|C_{tr}/T_{tr}| \sim
|(P_t V_{tb}^* V_{td}) / (T_{tr} V_{ub}^* V_{ud})| \sim 0.2$
\cite{SU3EWPs}. ($T_{tr}$ and $C_{tr}$ are the tree-level
color-allowed and color-suppressed amplitudes, respectively.) While
the $|P/T|$ ratio in Table~\ref{SMfit} is in line with these
expectations, $|C/T|$ is quite a bit larger than expected. We
therefore see that, although the SM can account for the $B\to\pi\pi$
data, it appears to give hadronic parameters which are at odds with
theoretical expectations. It is therefore worthwhile to re-examine the
fit, including the theoretical hadronic inputs of $|C/T|$ and $|P/T|$.

Consider first the ratio $|C/T|$:
\beq
\left\vert {C \over T} \right\vert = \left\vert { C_{tr} - (P_u - P_c)
\over T_{tr} + (P_u - P_c) } \right\vert ~.
\eeq
A numerical value of this ratio depends on estimates of the magnitudes
and strong phases of $C_{tr}/T_{tr}$ and $(P_u - P_c)/T_{tr}$. The
theoretical estimate of $|C_{tr}/T_{tr}|$ is $\sim 0.2$
\cite{SU3EWPs}. We add a theoretical uncertainty and take
$|C_{tr}/T_{tr}| = 0.2 \pm 0.1$. We also have a theoretical estimate
of $|(P_t V_{tb}^* V_{td}) / (T_{tr} V_{ub}^* V_{ud})|$
\cite{SU3EWPs}. Including an error, it is taken to be $0.2 \pm
0.1$. However, it is expected that $P_u$ and $P_c$ are somewhat
smaller than $P_t$. We assume no particular cancellations, and take
$|(P_u - P_c)/T_{tr}| = 0.1 \pm 0.1$.

For the strong phases, we proceed as follows. All strong phases are
due to rescattering from intermediate states \cite{rescatter}.
Consider first the SM diagram $P_c$ [Eq.~(\ref{Pdefs})]. Its strong
phase arises principally from the rescattering of the ${\bar b} \to
{\bar c} c {\bar d}$ tree diagram, $T_c$. That is, the $(V-A) \times
(V-A)$ $T_c$ (${\bar b} \to {\bar c} c {\bar d}$) rescatters into the
$(V-A) \times V$ $P_c$ (${\bar b} \to {\bar u} u {\bar d}$). The
Wilson coefficients in the effective Hamiltonian indicate that the
size of the rescattered $P_c$ amplitude is only about 5--10\% of that
of the $T_c$ amplitude. Thus, rescattering costs a factor of about
10--20. However, the Wilson coefficients also show that the $T_c$
diagram is about 10--20 times as big as the $P_c$ penguin
diagram. Thus, the rescattered $T_c$ contributes a strong phase of
$O(1)$ to $P_c$.

There is also a contribution to the strong phase of $P_c$ from
rescattering from states created by $P_c$ itself
(``self-rescattering''). However, since rescattering costs a factor of
10--20, the strong phase generated from this self-rescattering is much
smaller than that generated by rescattering from $T_c$. This effect is
therefore subdominant, and the strong phase from this
self-rescattering can be neglected.

The diagram $P_u$ [Eq.~(\ref{Pdefs})] also receives a large $O(1)$
strong phase due to rescattering from the ${\bar b} \to {\bar u} u
{\bar d}$ tree diagram, $T_u$. Self-rescattering of $P_u$ is
negligible.

We therefore see that the principal source of strong phases is from
rescattering from the tree diagrams $T_c$ and $T_u$. Thus, the strong
phases of $P_u$ and $P_c$ are large and can take essentially any
value. On the other hand, $P_t$ is essentially real and the strong
phase of $T_{tr}$ is very small (it is pure self-rescattering).
Finally, the strong phase of the diagram $C_{tr}$ comes from
rescattering from $T_c$ and is smaller than that of $P_u$/$P_c$, but
not negligible. We take it to be $0 \pm 0.2$.

The ratio $|P/T|$ is given by
\beq
\left\vert {P \over T} \right\vert = \left\vert { (P_t - P_c) V_{tb}^*
V_{td} \over [T_{tr} + (P_u - P_c)] V_{ub}^* V_{ud} } \right\vert ~.
\eeq
The same inputs as above are necessary for the estimate of this ratio.

In all cases, we assume that the magnitude ratios and strong phases
have Gaussian and flat distributions, respectively. Allowing all
quantities to vary in their allowed ranges according to their
(assumed) distributions, we find the following estimates:
\beq
r_{\sss C} = \left\vert {C \over T} \right\vert = 0.3 \pm 0.2 ~,\ \ \ \
r_{\sss P} = \left\vert {P \over T} \right\vert = 0.2 \pm 0.1 ~.
\label{theoryinput}
\eeq
These are the only two theoretical inputs of ratios of magnitudes of
amplitudes, and they are thought to be reasonably reliable. In
performing the fit in this section we exclude other inputs, which have
to do with estimates of the sizes of individual diagrams and/or strong
phases, and which are considered to be less solid. (Note that the
above procedure gives $\delta_{\sss C} - \delta_{\sss T} = (0 \pm
70)^\circ$. This is consistent with the result found in
Table~\ref{SMfit}.)

\begin{table}
\caption{Results of a fit within the SM. Inputs are the $B \to \pi\pi$
data (Table~\ref{T:data}), independent determinations of the CP
phases, and the theoretical expectations of
Eq.~(\ref{theoryinput}). All angles are in degrees.}
\begin{ruledtabular}
\begin{tabular}{ccccc}
$|T|$ & $|C|$ & $|P|$ & $\delta_{\sss C} - \delta_{\sss T}$ &
$\delta_{\sss P} - \delta_{\sss T}$ \\
\hline
$22.3 \pm 1.0$ & $16.7 \pm 2.0$ & $5.3 \pm 1.5$ & $-61.6 \pm 15.1$ &
$-51.7 \pm 20.1$ \\
\end{tabular}
\end{ruledtabular}
\label{SMfittwo} 
\end{table}

We now redo the SM fit, including the theoretical input of
Eq.~(\ref{theoryinput}) above. We again find a single solution for the
five unknown theoretical parameters (Table~\ref{SMfittwo}). The
results are similar to those found for the fit without theoretical
expectations (Table~\ref{SMfit}). However, in this case, we have a
much worse fit: $\chi^2_{min}/d.o.f.\ = 6.7/4$.

One explanation is statistical fluctuation. The solution is driven by
the unexpectedly large branching ratio of $B^0 \to \pi^0 \pi^0$. As
more data is taken, this branching ratio might decrease, giving a fit
which is more in line with theoretical expectations. (Note that
$\chi^2_{min}/d.o.f.\ = 6.7/4$ only corresponds to a deviation of
about $1.5\sigma$. Thus, we must stress that the statistical
discrepancy in the SM fit is not very strong at present.) A second
explanation is that the theoretical expectation of $|C/T|$ is
wrong. For example, this can happen if $P_u$ is larger than expected
\cite{largePu} (if $P_c$ is larger, the size of $P$ will in general be
affected). The point is that the naive estimates of $|C_{tr}/T_{tr}|
\sim |P_t/T_{tr}| \sim 0.2$ are for the amplitudes uncontaminated by
penguin contributions, whereas the $T$, $C$ and $P$ amplitudes which
appear in the fit all contain $P_u$ and/or $P_c$ contributions
[Eqs.~(\ref{Pdefs}), (\ref{mixeddiags})]. If $P_u$ is sizeable, and if
the strong phases are such that there is destructive interference
between the color-allowed tree amplitude $T_{tr}$ and $P_u - P_c$ (for
$T$) and/or constructive interference between the color-suppressed
tree amplitude $C_{tr}$ and $P_u - P_c$ (for $C$), one can account for
the large $|C/T|$ ratio. However, this solution is somewhat
fine-tuned, and we have no idea why $P_u$ should be larger than
expected.

A third explanation, which is the one investigated in this paper, is
that new physics is manifesting itself in the $B\to\pi\pi$ data.

\section{New Physics: No Hadronic Input}

As discussed in the introduction, the $B\to\pi\pi$ amplitudes can be
written in terms of isospin amplitudes or diagrams. In the SM these
are related as
\bea
A_2 = \frac{1}{3} (T + C) e^{i \gamma} ~,~~ && \bar A_2 = \frac{1}{3}
(T + C) e^{-i \gamma} ~, \nn\\
A_0 = \frac{1}{3} (2 T - C) e^{i \gamma} + P e^{-i \beta} ~,~~ && \bar
A_0 = \frac{1}{3} (2 T - C) e^{-i \gamma} + P e^{i \beta} ~,
\label{isoamps}
\eea
where the diagrams $T$, $C$, $P$ include strong phases, but their weak
phases have now been written explicitly. We therefore see that there
are only two areas in which new physics can enter: the $I=0$ or $I=2$
amplitudes. In the most general case, these NP contributions have
arbitrary phases, so that the isospin amplitudes can now be written
\bea
A_2 = \frac{1}{3} (T + C) e^{i \gamma} + N_2 ~,~~ && \bar A_2 =
\frac{1}{3} (T + C) e^{-i \gamma} + \bar N_2 ~, \nn\\
A_0 = \frac{1}{3} (2 T - C) e^{i \gamma} + P e^{-i \beta} + N_0 ~,~~
&& \bar A_0 = \frac{1}{3} (2 T - C) e^{-i \gamma} + P e^{i \beta} + \bar
N_0 ~,
\eea
where $N_0$, $\bar N_0$, $N_2$, and $\bar N_2$ are complex numbers.
Although the following discussion is general, it is assumed that the
NP effects, if present, are roughly the same size as the $B\to\pi\pi$
penguin amplitudes. There are several reasons for this. First, from
the theoretical point of view, it is most likely that the NP will
affect ${\bar b} \to {\bar d} q {\bar q}$ loop-level
processes. Second, if the NP were larger, of tree-level size, it is
probable that we would already have seen evidence for it, through
branching ratios, direct CP asymmetries, etc. Finally, if the NP
amplitudes are smaller, effects due to the SM EWP amplitudes become
important, and the NP will be more difficult 
(if not impossible) to detect.

Now, it was shown in Ref.~\cite{reparam} that any complex amplitude
can be written in terms of two weak phases as follows:
\beq
N = N_{\phi_{A1}} e^{i \phi_{A1}} + N_{\phi_{A2}} e^{i \phi_{A2}}
~~,~~~~ \bar N = N_{\phi_{A1}} e^{-i \phi_{A1}} + N_{\phi_{A2}} e^{-i
\phi_{A2}} ~,
\eeq
where
\beq
N_{\phi_{A1}} = \frac{N e^{-i \phi_{A2}} - \bar N e^{i \phi_{A2}} }{2
i \sin{(\phi_{A1} - \phi_{A2})}} ~~,~~~~
N_{\phi_{A2}} = \frac{N e^{-i \phi_{A1}} - \bar N e^{i \phi_{A1}} }{2
i \sin{(\phi_{A2} - \phi_{A1})}} ~.
\eeq
Note that the same complex numbers $N_{\phi_{A1}}$ and $N_{\phi_{A2}}$
appear in the expressions for $N$ and ${\bar N}$. 
This means that each includes a magnitude and a strong (CP-even)
phase,
but that the weak phases ($\phi_{A1}$ and $\phi_{A2}$, respectively)
have been taken out explicitly.

These relations, which are known as reparametrization invariance, can
be applied to $N_2$ and $N_0$. The key point is that any two weak
phases can be used. We therefore choose to express $N_0$ in terms of
$\gamma$ and $-\beta$; $N_2$ is written in terms of $\gamma$ and
another weak phase. The choice of this second weak phase is arbitrary;
the only condition is that it be known. For convenience, we take it to
be zero. Writing $N_2$ and $N_0$ in this way, one obtains \cite{BBLS}
\bea
A_2 = \frac{1}{3} (t + c) e^{i \gamma} + N_{2,0}\ ~,~~ && \bar A_2 =
\frac{1}{3} (t + c) e^{-i \gamma} + N_{2,0}\ ~, \nn\\
A_0 = \frac{1}{3} (2 t - c) e^{i \gamma} + p e^{-i \beta} ~,~~ && \bar
A_0 = \frac{1}{3} (2 t - c) e^{-i \gamma} + p e^{i \beta} ~,
\label{isoampsNP}
\eea
where
\bea
t + c &=& T + C + 3 N_{2,\gamma} ~, \nn\\
2 t - c &=& 2 T- C + 3 N_{0,\gamma} ~, \nn\\
p &=& P + N_{0,-\beta} ~,
\label{comb_tcp}
\eea
with
\bea
N_{2, \gamma} &=& i \frac{\bar N_2 - N_2}{2 \sin{\gamma}}, \nn\\
N_{2,0} &=& \frac{\bar N_2 + N_2}{2} - i \frac{\bar N_2 - N_2}{2
\tan{\gamma}}, \nn\\
N_{0,\gamma} &=& \frac{\bar N_0 + N_0}{2} \frac{\sin{\beta}}{\sin{(\beta
+ \gamma)}} + i \frac{\bar N_0 - N_0}{2}
\frac{\cos{\beta}}{\sin{(\beta + \gamma)}}, \nn \\
N_{0,-\beta} &=& \frac{\bar N_0 + N_0}{2}
\frac{\sin{\gamma}}{\sin{(\beta + \gamma)}} - i \frac{\bar N_0 -
N_0}{2} \frac{\cos{\gamma}}{\sin{(\beta + \gamma)}} ~.
\label{repara_nisto}
\eea

Comparing the expressions for the isospin amplitudes in
Eqs.~(\ref{isoamps}) and (\ref{isoampsNP}), one notices several things
\cite{BBLS}. First, the expressions for $A_0$ and ${\bar A}_0$ in
these two equations have the same form. This implies that one cannot
detect $I=0$ NP without hadronic input. Second, the expressions for
$A_2$ and $\bar A_2$ do not have the same form. Thus, even without
hadronic input, one can detect NP, but only the $N_{2,0}$ piece.
Because $N_{2,0}$ is a complex number, there are two tests for NP,
related to its magnitude and phase (assuming that $N_{2,0}$ is
nonzero). These are: a nonzero direct CP asymmetry $C_{+0}$ (if the
strong phase of $N_{2,0}$ is different from that of $t+c$) and a
difference between the value of $\gamma$ extracted from $B\to\pi\pi$
decays and that obtained from other, independent measurements.

Absorbing the NP parameters as above, we can now count the number of
parameters. There are nine: the four magnitudes of $t$, $c$, $p$ and
$N_{2,0}$, three relative strong phases, and the two weak phases
$\beta$ and $\gamma$. Thus, if the values of the weak phases are
taken from independent determinations, the seven $B\to\pi\pi$
observables are sufficient to perform a fit (this fit is taken from
Ref.~\cite{BBLS}.)

As in the SM fit, we assume that the values of weak CP phases are
already known. The present $B \to \pi \pi$ measurements are detailed
in Table~\ref{T:data}. In fitting to the data, we find four solutions,
shown in Table~\ref{T:fit_diagrammatic}. Although the second solution
gives a nonzero value of $|N_{2,0}|$, the other three solutions give
values which are consistent with zero. We therefore conclude that, if
one pays no attention to hadronic quantities, as done in this fit, the
present $B\to\pi\pi$ data exhibit no compelling signs of new physics.
This is to be expected since, as shown in Table~\ref{SMfit}, the SM
can account for the $B\to\pi\pi$ data if hadronic theoretical
expectations are ignored. That is, there is no need for NP.

\begin{table}
\caption{Results of a fit including NP, using the parametrization of
Eq.~(\ref{isoampsNP}). Inputs are the $B \to \pi\pi$ data
(Table~\ref{T:data}) and independent determinations of the CP phases.
We have factored out the (unphysical) overall phase $\delta_{N_{2,0}}
= \arg N_{2,0}$. The magnitudes are measured in $eV$ and the phases in
degrees.}
\begin{ruledtabular}
\begin{tabular}{ccccccc}
$|t|$  & $|c|$  &  $|p|$ & $|N_{2,0}|$ & 
              $\delta_t - \delta_{N_{2,0}}$ &
              $\delta_c - \delta_{N_{2,0}}$ &
              $\delta_p - \delta_{N_{2,0}}$ \\
\hline
     6.1 $\pm$ 2.7 & 9.9 $\pm$ 13.7 & 12.9 $\pm$ 3.2 & 9.6 $\pm$ 6.5 &
     81.5 $\pm$ 70.5 & $-$40.5 $\pm$ 90.0& 22.3 $\pm$ 74.1 \\
      2.8 $\pm$ 2.6 & 19.8 $\pm$ 23.8 & 11.4 $\pm$ 6.6 & 13.2 $\pm$
      1.3 & 41 $\pm$ 108 & $-$174 $\pm$ 9 & $-$48.6 $\pm$ 64.2\\
     22.8 $\pm$ 4.0 & 18.2 $\pm$ 6.7 & 7.3 $\pm$ 6.5 & 2.7 $\pm$ 9.3 &
     $-$156 $\pm$ 52 & 155 $\pm$ 32 & 157 $\pm$ 20 \\
     19.6 $\pm$ 3.9 & 6.1 $\pm$ 22.4 & 6.4 $\pm$ 1.7 & 6.0 $\pm$ 8.4 &
     $-$19.1 $\pm 43.9$ & 68.9 $\pm$ 174 & $-$127 $\pm$ 35 \\
\end{tabular}
\end{ruledtabular}
\label{T:fit_diagrammatic}
\end{table}

Note that we obtain $\chi^2_{min}/d.o.f.\ = 0.0049/0$, instead of zero
as expected. This occurs because the current data are slightly
inconsistent with the isospin $\{A_0, A_2\}$ description. That is, for
the central values
\beq
\cos(\delta_2 - \delta_0) = 
\frac{{2 \over 3} |A_{+0}|^2 + |A_{+-}|^2 -2 |A_{00}|^2}{
2 \sqrt{2} |A_{+0}| |A_0|}
= 1.07,
\eeq
where $|A_0|$ is given by
\beq
|A_0|^2 = \frac{2}{3}
\left( 
-\frac{2}{3} |A_{+0}|^2 + |A_{+-}|^2 + |A_{00}|^2
\right).
\eeq
Thus, our fit gives a nonzero value of $\chi^2_{min}/d.o.f.$

\section{New Physics: Hadronic Input}

In the previous section, we performed a fit without hadronic
input. However, as discussed earlier, we do have some theoretical
hadronic information which can be added to the fit. This is done in
this section.

\subsection{Preliminary remarks}

The equations in Eqs.~(\ref{isoampsNP})--(\ref{repara_nisto}) can be
inverted to give
\bea
p = P + N_{0,-\beta} &=&
\frac{\bar A_0\, e^{i \gamma} - A_0\, e^{-i \gamma}}{
2 i \sin{(\beta + \gamma)}},
\nn\\*[4mm]
t = T + (N_{2,\gamma} + N_{0,\gamma}) &=&
- \frac{\bar A_2  - A_2}{2 i \sin{\gamma}}
- \frac{\bar A_0\, e^{-i \beta} - A_0\, e^{i \beta}}{
2 i \sin{(\beta + \gamma)}},
\nn\\*[4mm]
c = C + (2 N_{2,\gamma} - N_{0,\gamma})
&=&
- 2\; \frac{\bar A_2  - A_2}{2 i \sin{\gamma}}
+ \frac{\bar A_0\, e^{-i \beta} - A_0\, e^{i \beta}}{
2 i \sin{(\beta + \gamma)}},
\nn\\*[4mm]
N_{2,0} &=& 
\frac{\bar A_2\, e^{i \gamma} - A_2\, e^{-i \gamma}}{
2 i \sin{\gamma}}.
\label{eq:analytic}
\eea

Now, in the preceding sections we introduced three equivalent sets of
variables. The set closest to experiment is $B_{+0}$, $B_{+-}$,
$B_{00}$, $C_{+0}$, $C_{+-}$, $C_{00}$, $S_{+-}$. From this we can
extract the isospin set $|A_0|$, $|\bar A_0|$, $|A_2|$, $|\bar A_2|$,
$\delta_2 - \bar \delta_0$, $\delta_0 - \bar \delta_0$, $\bar \delta_2
- \bar \delta_0$, or, equivalently --- \textit{c.f.\/}
Eqs.~(\ref{eq:analytic}) --- the diagrammatic set $|t|$, $|p|$, $|c|$,
$|N_{2,0}|$, $\delta_t - \delta_{N_{2,0}}$, $\delta_c -
\delta_{N_{2,0}}$, and $\delta_p - \delta_{N_{2,0}}$. These three
sets are completely determined from experiment.

Eqs.~(\ref{eq:analytic}) show us that if theoretical hadronic input is
added involving any combination of $T$, $C$, and/or $P$, we will be
able to probe new physics in $N_{2,\gamma}$ and/or the I=0 new physics
amplitudes. In particular, one sees that assumptions about $P$, $T$,
and $C$ will allow us to probe $N_{0,-\beta}$, $N_{2,\gamma} +
N_{0,\gamma}$, and $2 N_{2,\gamma} - N_{0,\gamma}$, respectively. For
example, given the experimental fit of $|p|$ and a theoretical
assumption about $|P|$, $N_{0,-\beta}$ is constrained by
\beq
|p|^2 = |P^2| + |N_{0, -\beta}|^2 + 2 |P|\, |N_{0, -\beta}|\,
 \cos{\theta_{\sss P}}
\eeq
where $\theta_{\sss P} = {\rm arg}(N_{0, -\beta} P^\ast)$. Thus, $|p| \neq
|P|$ is a sign of new physics in $N_{0,-\beta}$. Indeed,
\bea
\left| |p| - |P| \right| \leq
|N_{0,-\beta}| &\leq&
|p| + |P|,
\nn\\
\left| |t| - |T| \right| \leq
|N_{2,\gamma} + N_{0,\gamma}| &\leq&
|t| + |T|,
\nn\\
\left| |c| - |C| \right| \leq
|2 N_{2,\gamma} - N_{0,\gamma}| &\leq&
|c| + |C|.
\eea
We can also identify the combinations that probe $N_{2,\gamma}$ and
$N_{0,\gamma}$ independently
\bea
\left| |t + c| - |T + C| \right| \leq
3 |N_{2,\gamma}| &\leq&
|t + c| + |T + C|,
\nn\\
\left| |2 t - c| - |2 T - C| \right| \leq
3 |N_{0,\gamma}| &\leq&
|2 t - c| + |2 T - C|.
\eea
Recall that one cannot identify new physics in $I=0$ without hadronic
assumptions. But, as we have just shown, it takes only \textit{one
single} theoretical assumption (about $|P|$) in order to be able to
probe new physics in $I=0$.

Unfortunately, estimates of the sizes of individual diagrams are
thought to be less reliable than the two theoretical inputs of ratios
of magnitudes of amplitudes in Eq.~(\ref{theoryinput}). Using
Eqs.~(\ref{eq:analytic}) we find
\bea
\frac{P}{T} - \frac{p}{t}
&=&
- \frac{N_{0,-\beta}}{T} + \frac{N_{2,\gamma} + N_{0,\gamma}}{T}\,
\frac{p}{t},
\nn\\
\frac{C}{T} - \frac{c}{t}
&=&
- \frac{2\, N_{2,\gamma} - N_{0,\gamma}}{T} 
+ \frac{N_{2,\gamma} + N_{0,\gamma}}{T}\,
\frac{c}{t},
\label{qui_1}
\eea
leading to
\bea
\left|
- \frac{N_{0,-\beta}}{T} + \frac{N_{2,\gamma} + N_{0,\gamma}}{T}
\,\frac{p}{t}
\right|
&\geq&
\left|
\left|\frac{P}{T}\right| - \left|\frac{p}{t}\right|
\right|,
\nn\\
\left|
- \frac{2\, N_{2,\gamma} - N_{0,\gamma}}{T} 
+ \frac{N_{2,\gamma} + N_{0,\gamma}}{T}\,
\frac{c}{t}
\right|
&\geq&
\left|
\left|\frac{C}{T}\right| - \left|\frac{c}{t}\right|
\right|.
\label{qui_2}
\eea
These equations show which combinations of new physics parameters are
probed when the experimental observable $|p/t|$ ($|c/t|$) differs from
the SM theoretical prediction for $|P/T|$ ($|C/T|$, respectively).
Sadly, these combinations are rather complicated and, in particular,
we cannot disentangle $N_{2,\gamma}$ from the $I=0$ new physics
contributions using exclusively theoretical predictions for the SM
$|P/T|$ and $|C/T|$.

\subsection{Fitting for the new-physics parameters}

We will now consider several scenarios for the NP, attempting to fit
for all free parameters available within each scenario. If we assume
that both $I=0$ and $I=2$ NP are present, then there are 13 unknown
parameters: the seven magnitudes of $T$, $C$, $P$, $N_{0,\gamma}$,
$N_{0,-\beta}$, $N_{2,\gamma}$, $N_{2,0}$ and the six relative phases
(recall that we assume that the SM CP phases are known). Given the
seven observables, a complete fit would require six theoretical
hadronic inputs, which is unmanageable. For this reason, we assume
that one type of NP dominates over the other, and consider $I=2$ and
$I=0$ NP individually.

\subsubsection{New physics exclusively in $I=2$}

Consider only $I=2$ NP. There are nine unknown parameters: five
magnitudes of $T$, $C$, $P$, $N_{2,\gamma}$ and $N_{2,0}$, and the
four relative phases. With seven $B\to\pi\pi$ observables, two
additional theoretical hadronic constraints are necessary in order to
measure all the $I=2$ NP parameters, which we take to be those in
Eq.~(\ref{theoryinput}). 

Before performing this fit, we can make some general observations.
Recall that $N_{2,0}$ is determined from experiment. In the absence
of $I=0$ new physics, Eqs.~(\ref{comb_tcp}) become
\bea
t + c &=& T + C + 3 N_{2,\gamma} ~, \nn\\
2 t - c &=& 2 T- C  ~, \nn\\
p &=& P  ~,
\label{comb_tcp_2}
\eea
which, combined with our estimates for $r_{\sss C}$ and $r_{\sss P}$, allow us to
find explicit analytical formulas for $T$, $C$, $P$, and
$N_{2,\gamma}$. We do not include them, since they are not very
instructive, but they do provide a few interesting bounds,
\bea
\left| \frac{N_{2,\gamma}}{T} \right| &\geq& \left| 1 -
\frac{r_{\sss P}}{\left|p/t\right|}\right|, \nn\\
\left| \frac{N_{2,\gamma}}{T} \right| &\geq& \left| \frac{r_{\sss C} -
\left|c/t\right|}{ 2 - c/t} \right|.
\eea
Moreover, dividing the last two of Eqs.~(\ref{comb_tcp_2}), we find
\beq
r_{\sss C}\, e^{i(\delta_{\sss C} - \delta_{\sss T})}
= 2 - \left( \frac{2t-c}{p}\right)
r_{\sss P}\, e^{i(\delta_{\sss P} - \delta_{\sss T})},
\eeq
from which
\beq
\left| 2 - \left| \frac{2t-c}{p} \right| r_{\sss P} \right|
\leq r_{\sss C} \leq 2 + \left| \frac{2t-c}{p} \right| r_{\sss P}.
\label{sb_bound}
\eeq
Given the experimental determination of $t$, $c$, and $p$, this
constrains the choices of $(r_{\sss C}, r_{\sss P})$ which are
consistent with the hypothesis that all new physics appears
exclusively in $I=2$.

\begin{table}
\caption{Results of a fit assuming only $I=2$ NP, using the
parametrization of Eq.~(\ref{isoampsNP}). Inputs are the $B \to
\pi\pi$ data (Table~\ref{T:data}), independent determinations of the
CP phases, and the hadronic estimates of Eq.~(\ref{theoryinput}).  All
angles are in degrees.}
\begin{ruledtabular}
\begin{tabular}{ccccc}
$|T|$ & $|C|$ & $|P|$ & $|N_{2,\gamma}|$ & $|N_{2,0}|$ \\
$\delta_{\sss T} - \delta_{N_{2,0}} $ & $\delta_{\sss C} - \delta_{N_{2,0}}$ &
$\delta_{\sss P} - \delta_{N_{2,0}}$ & 
$\delta_{N_{2,\gamma}} - \delta_{N_{2,0}}$ & \\
\hline
$24.0 \pm 3.8$ & $7.6 \pm 5.6$ & $5.5 \pm 1.7$ & $10.3 \pm 11.4$ & 
$0.55 \pm 2.81$ \\
$-116 \pm 72$ & $-116 \pm 195$ & $163 \pm 69$ & $131 \pm 65$ & \\
\end{tabular}
\end{ruledtabular}
\label{NPfithad-2} 
\end{table}
%

We now redo the NP fit of Table~\ref{T:fit_diagrammatic} with the
theoretical hadronic input. Table~\ref{NPfithad-2} shows the result of
this fit. We obtain $\chi^2_{min}/d.o.f.\ = 0.27/0$. Note that we
expect $\chi^2_{min} = 0$ for zero degrees of freedom (i.e.\ a
solution to the equations). The fact that we don't obtain this
indicates that the equations to be solved are inconsistent, and
suggests a poor fit. However, at this point we are not certain of how
poor this fit is since we cannot test the goodness of fit with zero
degrees of freedom.

Fortunately, there is an alternative: consider again the four
solutions of Table~\ref{T:fit_diagrammatic}, which correspond to the
inclusion of NP without theoretical hadronic input. With hadronic
input, Eq.~(\ref{sb_bound}) should be satisfied. However, the four
solutions in Table~\ref{T:fit_diagrammatic} lead to
\beq
\left\vert 2t -c \over p \right\vert = 1.50 \pm 0.95,\; 2.17 \pm
3.30,\; 5.00 \pm 6.55,\; 6.20 \pm 1.41 ~,
\eeq
respectively. Using Eq.~(\ref{sb_bound}), we see that none of the
central values is consistent with $(r_{\sss C}, r_{\sss P}) = (0.3,
0.2)$. If we relied exclusively on the central values, this would show
explicitly that none of the solutions in
Table~\ref{T:fit_diagrammatic} actually provides a good fit, assuming
our theoretical input is correct. However, the errors are still quite
large, so that no such conclusion can be drawn.

There is another way to test the goodness-of-fit: we calculate
\beq
\cos(\delta_{\sss T} -\delta_{\sss C}) = \frac{4 + r_{\sss C}^2}{4
  r_{\sss C}} - \frac{r_{\sss P}^2}{4 r_{\sss C}} \left\vert 2 t -c
  \over p\right\vert ^2.
\label{cosprobs}
\eeq
Using the values of $t$, $c$ and $p$ given in
Table~\ref{T:fit_diagrammatic}, we find
\beq
\cos(\delta_{\sss T} -\delta_{\sss C}) = \{ 3.33 \pm 2.13,\; 3.25 \pm
 2.13,\; 2.57 \pm 2.84,\; 2.10 \pm 1.93 \} ~.
\eeq
All central values are much larger than 1, suggesting a poor fit.
However, when we take into account the error, the third value is
(just) consistent with $\cos(\delta_{\sss T} -\delta_{\sss C}) \le 1$,
so that this value is acceptable. (Note that the third solution also
has the smallest $\chi^2$.)

We therefore conclude that, with this parametrization of the new
physics, in which the $I=2$ NP has zero degrees of freedom, this type
of NP is still permitted, although the fit is not perfect. There may
be a small deviation of about $1\sigma$. This shows that it is still
possible to keep $(r_{\sss C}, r_{\sss P}) = (0.3, 0.2)$ by adding
$I=2$ NP.

Note that it is possible to improve the fit by choosing a particular
value of the NP weak phase. For instance, we could take only
$N_{2,\gamma}$ or $N_{2,0}$ to be nonzero. In both of these cases, the
number of degrees of freedom is increased to 2, and we obtain
acceptable fits. However, we prefer not to make any assumptions about
the nature of the NP (i.e.\ its weak phase), so we do not consider
this possibility here.

\subsubsection{New physics exclusively in $I=0$}

The case of $I=0$ NP is more complicated. Here there are the same
number of parameters and constraints --- nine --- but the effect of
the $B\to\pi\pi$ observables is different. First, if there is no $I =
2$ NP, the direct CP asymmetry in charged $B$ decays is automatically
zero, i.e.\ $C_{+0} = 0$. This means that the measurements of $BR(B^+
\to \pi^+ \pi^0)$ and $BR(B^- \to \pi^- \pi^0)$ do not provide
constraints on the NP. Instead, they give two different measurements
of the same quantity, $|T + C|$. In addition, in the presence of only
$I = 0$ NP, the situation is analogous to that of the SM: the $I=2$
amplitude has a well-defined weak phase, while the $I=0$ amplitude
does not. In the SM, the 6 $B\to\pi\pi$ measurements are sufficient to
obtain the various parameters, including the CP phase $\alpha$. That
is, some combination of $B\to\pi\pi$ measurements is equal to
$\alpha$. However, in our analysis, $\alpha$ is assumed to be known
independently. Therefore the combination of $B\to\pi\pi$ measurements
which is equal to $\alpha$ does not provide additional information.

Thus, for the case in which NP appears exclusively in $I=0$, we have
only five independent $B\to\pi\pi$ constraints, so that we need {\it
four} additional theoretical inputs to measure \textit{all} the $I=0$
NP parameters. Since we have only the two theoretical inputs of
Eq.~(\ref{theoryinput}), we conclude that it is not possible to
measure the $I=0$ NP parameters.

We note that one can fit for the $I=0$ NP parameters if one chooses
only one of the two terms $N_{0,\gamma}$ or $N_{0,-\beta}$, in which
case there are an equal number of parameters and constraints. However,
as was the case with $I=2$ NP, we prefer not to fix the form of the NP
at this time.

Notice that the quantities $t$, $c$ and $p$ can still be determined in
this case. The key point is that these are {\it observables},
derivable from the $B\to\pi\pi$ measurements
[Eqs.~(\ref{isoampsNP})--(\ref{repara_nisto})]. In fact, since we are
setting $N_{2,0}=0$ here, the values of $t$, $c$ and $p$ can be
determined from the SM fit in Table~\ref{SMfit}, with the replacements
$T \to t$, $C \to c$, and $P \to p$. What we cannot do is to
disentangle all the parameters $T$, $C$, $P$, $N_{0,\gamma}$, and
$N_{0,- \beta}$ from our knowledge of $t$, $c$ and $p$.

Fortunately, although in this case it is impossible to fit all
parameters, we may still gain some information about the new physics.
In the absence of $I=2$ new physics, Eqs.~(\ref{comb_tcp}) become
\bea
t + c &=& T + C ~, \nn\\
2 t - c &=& 2 T- C + 3 N_{0,\gamma} ~, \nn\\
p &=& P + N_{0,-\beta},
\label{comb_tcp_0}
\eea
while Eqs.~(\ref{qui_1}) become
\bea
\frac{P}{T} - \frac{p}{t}
&=&
- \frac{N_{0,-\beta}}{T} + \frac{N_{0,\gamma}}{T}\,
\frac{p}{t},
\nn\\
\frac{C}{T} - \frac{c}{t}
&=&
\frac{N_{0,\gamma}}{T} \left( 1 + \frac{c}{t} \right).
\label{qui_1_0}
\eea
The last equation leads to
\beq
\left| \frac{r_{\sss C} - \left|c/t\right|}{ 1 + c/t} \right| \leq \left|
\frac{N_{0,\gamma}}{T}\right| \leq \frac{r_{\sss C} + \left|c/t\right|}{
\left|1 + c/t \right|},
\label{qui_1_0g}
\eeq
meaning that, if we are lucky, a hadronic input on $r_{\sss C}$ might be
enough to detect the presence of $N_{0,\gamma}$ new physics. Using
the central values for $r_{\sss C}$ and for $t$, $c$, and $\delta_c -
\delta_t$ from Table~\ref{SMfit}, we find
\beq
0.35 \leq \left| \frac{N_{0,\gamma}}{T}\right| \leq 0.74\ .
\label{bound_I0}
\eeq
This means that, assuming the scenario in which new physics appears
exclusively in the $I=0$ channel and that $r_{\sss C} \sim 0.3$, the
conclusion is that there might be evidence for new physics in
$N_{0,\gamma}$.

Given the bound on $|N_{0,\gamma}/T|$ in Eq.~(\ref{qui_1_0g}), one may
check whether the magnitudes $r_{\sss P}$, $|p/t|$, and $\left|
(N_{0,\gamma}/T)\, (p/t) \right|$ can close a triangle. If they
cannot, then the first of Eq.~(\ref{qui_1_0}) can be used to place a
lower bound on $|N_{0,-\beta}/T|$, thus detecting new physics in
$N_{0,-\beta}$. We conclude that, given appropriate experimental
numbers and the two hadronic inputs $r_{\sss C}$ and $r_{\sss P}$, one
might be able to get information on both $|N_{0,\gamma}/T|$ and
$|N_{0,-\beta}/T|$. However, the current experimental data, which
implies that $|N_{0,\gamma}| \neq 0$ in this scenario ---
\textit{c.f.\/} Eq.~(\ref{bound_I0}) --- is consistent with
$N_{0,-\beta}=0$.

Summarizing, we see that the theory input $(r_{\sss C}, r_{\sss P})
\sim (0.3, 0.2)$, along with the current experimental results, can be
accommodated by NP in the $I=0$ channel. (Note that the errors are
still very large, so that new physics is not {\it required}.)

\section{New Physics: Strong Phases}

In the previous section, we saw that, with theoretical hadronic input,
$I=2$ new physics gave a poorish fit, and one could not even fit for
all the $I=0$ NP parameters. In both cases, the difficulties could be
traced to the large number of theoretical parameters. This then begs
the question: is it possible to reduce the number of parameters?  As
we explore in this section, the answer is yes, but it requires using a
different parametrization to describe the new physics.

In the discussion surrounding the SM fit, we pointed out that strong
phases are due to rescattering. This has implications for the strong
phases of the NP amplitudes. The strong phase for a particular NP
operator can arise only due to rescattering from itself or another NP
operator. In either case, we have self-rescattering. However, it is
reasonable to expect that the strong phases generated by such
rescattering are not $O(1)$, but only about 5--10\%. That is, the NP
strong phases are expected to be small \cite{BNP}. Since we can
multiply any amplitude by an arbitrary phase, a more accurate
statement is that all NP amplitudes (both $I=2$ and $I=0$) are
expected to have the same strong phase. The theoretical error incurred
by this assumption is expected to be at the level of 5--10\%.

Let us illustrate these features by considering $I=0$, for example.
The $I=0$ NP amplitude is
\beq
N_0 = |N_{0,a}| \, e^{i \Phi^{\sss NP}_{0,a}} \, e^{i \delta^{\sss
NP}_{0,a}} + |N_{0,b}| \, e^{i \Phi^{\sss NP}_{0,b}} \, e^{i \delta^{\sss
NP}_{0,b}} + ...,
\eeq
where ``$a$, $b$, ...'' labels the various possible contributions.
Within the above model, the strong phases for all contributions to a
particular NP amplitude are approximately equal. In this case, one can
combine all contributions into a single effective NP amplitude, with
well-defined weak and strong phases:
\beq
N_0 \to |N_{0,eff}| \, e^{i \Phi^{\sss NP}_{0,eff}} \,
e^{i \delta^{\sss NP}} ~.
\label{david}
\eeq
Here, $\Phi^{\sss NP}_{0,eff}$ is a weak phase which changes sign in
the CP-conjugate amplitudes; $\delta^{\sss NP}$ is the strong
phase. The $I=2$ NP amplitude has a similar expression with the same
NP strong phase.

It is also possible to rewrite the $I=0$ and $I=2$ amplitudes using
reparametrization invariance with the specific choices in
Eqs.~(\ref{repara_nisto}), but this is not terribly illuminating, so
we will stick to Eq.~(\ref{david}).

With this form for the NP amplitude, we write
\bea
-\sqrt{2} A(B^+ \to \pi^+ \pi^0) & = & (T + C) e^{i\gamma} + 3
 \left\vert N_{2,eff} \right\vert \, e^{i \Phi^{\sss NP}_{2,eff}} e^{i
 \delta^{\sss NP}} ~, \nn\\ 
- A(\bd \to \pi^+ \pi^-) & = & T e^{i\gamma} + P e^{-i\beta} +
 \left\vert N_{2,eff} \right\vert \, e^{i \Phi^{\sss NP}_{2,eff}} e^{i
 \delta^{\sss NP}} + \left\vert N_{0,eff} \right\vert \, e^{i
 \Phi^{\sss NP}_{0,eff}} e^{i \delta^{\sss NP}} ~, \nn\\
-\sqrt{2} A(\bd \to \pi^0 \pi^0) & = & C e^{i\gamma} - P e^{-i\beta} +
 2 \left\vert N_{2,eff} \right\vert \, e^{i \Phi^{\sss NP}_{2,eff}}
 e^{i \delta^{\sss NP}} - \left\vert N_{0,eff} \right\vert \, e^{i
 \Phi^{\sss NP}_{0,eff}} e^{i \delta^{\sss NP}} ~.
\label{newNPparam}
\eea
The weak phases change sign in the $\bar B$ decay amplitudes. Note
that, with this parametrization, $I=2$ and $I=0$ NP is described by a
single amplitude each, instead of two as was done in the
parametrization of reparametrization invariance.

As before, we first assume that both $I=0$ and $I=2$ NP are
present. In this case, there are 11 unknown parameters: the five
magnitudes $|T|$, $|C|$, $|P|$, $\left\vert N_{0,eff} \right\vert $
and $\left\vert N_{2,eff} \right\vert $, four relative strong phases,
and the two NP weak phases $\Phi^{\sss NP}_{0,eff}$ and $\Phi^{\sss
NP}_{2,eff}$. Since there are only 9 constraints (7 $B\to\pi\pi$
measurements, along with two theoretical hadronic inputs), there are
more theoretical parameters than there are observables, and we cannot
perform a fit. We therefore assume that one NP amplitude is dominant.

\subsubsection{New physics exclusively in $I=2$}

For the case in which one has only $I=2$ NP, there are 8 theoretical
parameters and 9 measurements/constraints. We can therefore perform a
fit, whose results are shown in Table~\ref{I=2NPfit}. In this case, we
find a single solution, with $\chi^2_{\rm min}/d.o.f.\ = 0.298/1$. Note
that this fit includes the same difficulties with $\cos(\delta_{\sss
T} -\delta_{\sss C})$ detailed in Eq.~(\ref{cosprobs}). However, since
there is one degree of freedom in this case, we obtain a good fit.

%
\begin{table}
\caption{Results of a fit assuming only $I=2$ NP, using the
parametrization of Eq.~(\ref{newNPparam}). Inputs are the $B \to
\pi\pi$ data (Table~\ref{T:data}), independent determinations of the
CP phases, and the hadronic estimates of Eq.~(\ref{theoryinput}).  All
angles are in degrees. We find $\chi^2_{\rm min}/d.o.f.\ = 0.298/1$.}
\begin{ruledtabular}
\begin{tabular}{cccc}
$|T|$ & $|C|$ & $|P|$ & $|N_{2,eff}|$ \\
$\delta_T - \delta^{\sss NP}$ & 
$\delta_C - \delta^{\sss NP}$ &
$\delta_P - \delta^{\sss NP}$ &
$\Phi^{\sss NP}_{2,eff}$  \\
\hline
$ 24.1 \pm 3.8$ & $7.5 \pm 5.6$ & $5.3 \pm 1.4$ & $10.3 \pm 13.2$ \\
$114 \pm 18$ & $115 \pm 189$ & $33.6 \pm 42.8$ & $59.0 \pm 5.5$  \\
\end{tabular}
\end{ruledtabular}
\label{I=2NPfit}
\end{table}
%

\subsubsection{New physics exclusively in $I=0$}

For the case with only $I=0$ new physics, we still cannot perform a
fit. As with $I=2$ NP, there are 8 theoretical parameters and 9
measurements/constraints. However, as discussed earlier, two of the
$B\to\pi\pi$ measurements are redundant and do not probe the NP
parameters. That is, for the purpose of a fit, there are effectively
only 7 measurements/constraints. We can therefore not measure the NP
parameters in this case.

\section{New Physics: More Hadronic Input}

On the previous section, we saw that if we assume that the NP
amplitudes are described by a single term each, we can fit for the
$I=2$ NP, but not for $I=0$. In the latter case, we need an additional
theoretical hadronic constraint.

This comes from the strong phases. We have argued that the strong
phase of $T$ is also due to self-rescattering, and should therefore be
equal to that of NP. We therefore take as the third input
\beq
\delta_{\sss T} - \delta^{\sss NP} = (0 \pm 11)^\circ,
\label{deltainput}
\eeq
where $I=0,2$. With this third input, we can perform a fit for all NP
quantities, in particular the $I=0$ NP parameters.

In this section we perform fits to both $I=2$ and $I=0$ NP. In order
to obtain the $I=0$ NP parameters, we need all three theoretical
inputs. For $I=2$ NP, we need only one, but for simplicity we use all
three.

For $I=2$ NP, the fit yields one solution for $|N_{2,eff}|$
(Table~\ref{NPsolutionsI=2}). This solution is rather intriguing, as
it yields a good value of $\chi^2_{min}/d.o.f.$ and a nonzero value of
$|N_{2,eff}|$ at more than $3\sigma$. However, we must emphasize that
this conclusion must be viewed with a great deal of skepticism since
this fit has much theoretical input. Indeed, though we did our best to
estimate the theoretical quantities, we could easily have incorrect
central values and/or errors. Still, the possibility of NP in
$B\to\pi\pi$ decays is rather interesting. As we noted in the
discussion of the SM (Secs.~2 and 3), the large value of $|C/T|$ in
the SM fit may be pointing to the presence of new physics.
Furthermore, the latest measurements of $B\to\pi K$ seem to indicate
the presence of NP in that sector \cite{BKpi}. The two NP solutions
may be related.

%
\begin{table}
\caption{Results of a fit assuming only $I=2$ NP, using the
parametrization of Eq.~(\ref{newNPparam}). Inputs are the $B \to
\pi\pi$ data (Table~\ref{T:data}), independent determinations of the
CP phases, and the hadronic estimates of Eqs.~(\ref{theoryinput}) and
(\ref{deltainput}). We find $\chi^2_{\sss min}/d.o.f.=0.66/2$.}
\begin{ruledtabular}
\begin{tabular}{cccc}
$|T|$ & $|C|$ & $|P|$ & $|N_{2,eff}|$ \\
$\delta_{\sss T} - \delta^{\sss NP}$ & $\delta_{\sss C} - \delta^{\sss
NP}$ & $\delta_{\sss P} - \delta^{\sss NP}$ & $\Phi^{\sss NP}_{2,eff}$
\\
\hline
$24.8 \pm 4.1$ & $8.1 \pm 5.8$ & $5.9 \pm 1.3$ & $7.0 \pm 2.0$ \\
$-3.7 \pm 7.5$ & $10.5 \pm 23.7$ & $-131 \pm 17$ & $58.9 \pm 5.7$ \\
\end{tabular}
\end{ruledtabular}
\label{NPsolutionsI=2}
\end{table}
%

In the fit for $I=0$ NP, there are two solutions for $|N_{0,eff}|$
(Table~\ref{NPsolutionsI=0}). In this case, because the number of
degrees of freedom is two, both fits are acceptable.

%
\begin{table}
\caption{Results of a fit assuming only $I=0$ NP, using the
parametrization of Eq.~(\ref{newNPparam}). Inputs are the $B \to
\pi\pi$ data (Table~\ref{T:data}), independent determinations of the
CP phases, and the hadronic estimates of Eqs.~(\ref{theoryinput}) and
(\ref{deltainput}). We find $\chi^2_{\sss min}/d.o.f.= 0.84/2$.}
\begin{ruledtabular}
\begin{tabular}{cccc}
$|T|$ & $|C|$ & $|P|$ & $|N_{0,eff}|$ \\
$\delta_{\sss T} - \delta^{\sss NP}$ & $\delta_{\sss C} - \delta^{\sss
NP}$ & $\delta_{\sss P} - \delta^{\sss NP}$ & $\Phi^{\sss NP}_{0,eff}$
\\
\hline
$40.0 \pm 18.4$ & $16.4 \pm 9.3$ & $8.0 \pm 5.4$ & $20.4 \pm 17.8$ \\
$4.6 \pm 9.9$ & $-118 \pm 54.5$ & $-28.9 \pm 28.9$ & $-86.6 \pm 23.1$
\\
\hline
$40.0 \pm 18.4$ & $16.4 \pm 9.3$ & $8.0 \pm 5.4$ & $20.4 \pm 17.8$ \\
$4.6 \pm 9.9$ & $-118 \pm 54.5$ & $-28.9 \pm 28.9$ & $-128 \pm 18.7$
\\
\end{tabular}
\end{ruledtabular}
\label{NPsolutionsI=0}
\end{table}
%

\section{Conclusions}

If there is new physics (NP) in $B\to\pi\pi$ decays, it can affect the
isospin $I=0$ or $I=2$ modes. In this paper, we have discussed several
ways of measuring these NP parameters. If no hadronic input is used,
one can never detect $I=0$ NP, and only one piece of the $I=2$ NP can
be observed. Thus, in order to fully detect and measure the NP, it is
necessary to add theoretical hadronic input.

The most obvious input consists of theoretical evaluations of the
amplitude ratios $|C/T|$ and $|P/T|$, where $T$ and $C$ are the
color-allowed and color-suppressed tree amplitudes, respectively, and
$P$ is the penguin contribution. (Note that the $C$ and $T$ amplitudes
that appear in the $B\to\pi\pi$ also contain penguin contributions.)
Given this input, there is enough information in principle to measure
the $I=2$ NP parameters, but not those for $I=0$. In practice, present
data yields a fit for $I=2$ NP which is not great, so that, with this
parametrization of new physics, in which the $I=2$ NP has zero degrees
of freedom, this type of NP is disfavored (but not ruled out). For
$I=0$ NP, although one cannot perform a fit to measure its parameters,
one can compute limits to detect its presence. The current data
suggest that this type of NP might be present, but the errors are
still very large.

One can then make a well-motivated assumption that the NP strong
phases are all about the same size. This leads to a new
parametrization of the NP with more degrees of freedom. In this case,
the addition of $I=2$ NP yields a good fit (although the presence of
NP is not favored), but one can still not fit for all the $I=0$ NP
parameters.

Finally, it is then possible to add a third reasonable assumption,
that the strong phases of $T$ and the NP are roughly equal. In this
case, one can again fit for $I=2$ NP, obtaining a good fit. More
importantly, there are now enough constraints to measure the $I=0$ NP
parameters, and one obtains a good fit in this case as well.

We must stress that there is no compelling evidence for NP in the
current $B\to\pi\pi$ data -- the SM can account for it with only a
$1.5\sigma$ discrepancy. However, should statistically-significant
evidence for NP be found in $B\to\pi\pi$ decays, the various methods
discussed in this paper will allow us to distinguish between $I=2$ and
$I=0$ NP, and measure its parameters.

\begin{acknowledgments}
S.B. and D.L. thank G. Azuelos, A. Datta and M. Gronau for helpful
discussions, while F.J.B. and J.P.S. thank L. Wolfenstein.  The work
of S.B. and D.L. is supported by NSERC of Canada.  F.J.B.  is
partially supported by the spanish M.E.C. under FPA2002-00612 and
HP2003-0079 (``Accion Integrada hispano-portuguesa'').  J.P.S. is
supported in part by the Portuguese \textit{Funda\c{c}\~{a}o para a
Ci\^{e}ncia e a Tecnologia} (FCT) under the contract CFTP-Plurianual
(777), and through the project POCTI/37449/FNU/2001, approved by the
Portuguese FCT and POCTI, and co-funded by FEDER.
\end{acknowledgments}



\begin{thebibliography}{99}

\bibitem{pdg} L. Wolfenstein, \prl{51}{1983}{1945}; S.~Eidelman {\it
et al.}  [Particle Data Group Collaboration], Phys.\ Lett.\ B {\bf
592} (2004) 1.

\bibitem{penguins} D.~London and R.~D.~Peccei, Phys.\ Lett.\ B {\bf
223}, 257 (1989); M.~Gronau, Phys.\ Rev.\ Lett.\ {\bf 63}, 1451
(1989), Phys.\ Lett.\ B {\bf 233}, 479 (1989); B.~Grinstein, Phys.\
Lett.\ B {\bf 229}, 280 (1989).

\bibitem{isospin} M.~Gronau and D.~London, Phys.\ Rev.\ Lett.\ {\bf
65}, 3381 (1990).

\bibitem{footnote1} In the SM, because EWP contributions are
negligible, we expect that the direct CP asymmetry in the charged $B$
decay vanishes, i.e.\ $|A(B^+ \to \pi^+ \pi^0)| = |A(B^- \to \pi^-
\pi^0)|$.

\bibitem{charles} J.~Charles, Phys.\ Rev.\ D {\bf 59}, 054007 (1999).

\bibitem{BBLS} S.~Baek, F.~J.~Botella, D.~London and J.~P.~Silva,
Phys.\ Rev.\ D {\bf 72}, 036004 (2005).

\bibitem{sin2beta} K.~Abe {\it et al.}  [BELLE Collaboration],
hep-ex/0408111; B.~Aubert {\it et al.}  [BABAR Collaboration],
hep-ex/0408127.

\bibitem{BDK} M.~Gronau and D.~Wyler, Phys.\ Lett.\ B {\bf 265}, 172
(1991); D.~Atwood, I.~Dunietz and A.~Soni, Phys.\ Rev.\ Lett.\ {\bf
78}, 3257 (1997). See also M.~Gronau and D.~London., Phys.\ Lett.\ B
{\bf 253}, 483 (1991); I.~Dunietz, Phys.\ Lett.\ B {\bf 270}, 75
(1991); N.~Sinha and R.~Sinha, Phys.\ Rev.\ Lett.\ {\bf 80}, 3706
(1998).

\bibitem{CKMfitter} The CKMfitter group, http://www.slac.stanford.edu \\
\null~~~~/xorg/ckmfitter/ckm\_results\_winter2005.html.

\bibitem{BRs} $B\to\pi\pi$ $BR$'s: A.~Bornheim {\it et al.}  [CLEO
Collaboration], Phys.\ Rev.\ D {\bf 68}, 052002 (2003); B.~Aubert {\it
et al.}  [BABAR Collaboration], arXiv:hep-ex/0408081 ($\pi^+ \pi^0$,
$\pi^0 \pi^0$), Phys.\ Rev.\ Lett.\ {\bf 89}, 281802 (2002) ($\pi^+
\pi^-$); Y.~Chao {\it et al.}  [Belle Collaboration], Phys.\ Rev.\ D
{\bf 69}, 111102 (2004) ($\pi^+ \pi^0$, $\pi^+ \pi^-$), K.~Abe {\it et
al.}  [Belle Collaboration], arXiv:hep-ex/0408101 ($\pi^0 \pi^0$).

\bibitem{ACPs} $A_{\sss CP}$ for $B\to\pi^+\pi^0$ and
$\bd\to\pi^0\pi^0$: B.~Aubert {\it et al.}  [BABAR Collaboration],
arXiv:hep-ex/0408081; Y.~Chao {\it et al.}  [Belle Collaboration],
Phys.\ Rev.\ Lett.\ {\bf 93}, 191802 (2004) ($\pi^+ \pi^0$), K.~Abe
{\it et al.}  [Belle Collaboration], arXiv:hep-ex/0408101 ($\pi^0
\pi^0$).

\bibitem{pi+pi-} $A_{\sss CP}$ and $S_{\sss CP}$ for
$\bd\to\pi^+\pi^-$: B.~Aubert {\it et al.}  [BABAR Collaboration],
arXiv:hep-ex/0408089; K.~Abe {\it et al.}  [Belle Collaboration],
arXiv:hep-ex/0502035.
  
\bibitem{HFAG} The Heavy Flavor Averaging Group, \\
http://www.slac.stanford.edu/xorg/hfag/

\bibitem{SU3EWPs} For example, see M.~Gronau, O.~F.~Hernandez,
D.~London and J.~L.~Rosner, Phys.\ Rev.\ D {\bf 52}, 6374 (1995).

\bibitem{rescatter} For example, see L. Wolfenstein and F. Wu,
arXiv:hep-ph/0506224, and references therein.

\bibitem{largePu} For example, see A.~J.~Buras, R.~Fleischer,
S.~Recksiegel and F.~Schwab, Nucl.\ Phys.\ B {\bf 697}, 133 (2004);
C.~W.~Chiang, M.~Gronau, J.~L.~Rosner and D.~A.~Suprun, Phys.\ Rev.\ D
{\bf 70}, 034020 (2004); H.~Y.~Cheng, C.~K.~Chua and A.~Soni, Phys.\
Rev.\ D {\bf 71}, 014030 (2005); L. Wolfenstein and F. Wu,
Ref.~\cite{rescatter}; Y.~Grossman, A.~Hocker, Z.~Ligeti and
D.~Pirjol, arXiv:hep-ph/0506228.

\bibitem{reparam} F.~J.~Botella and J.~P.~Silva, Phys.\ Rev.\ D {\bf
71}, 094008 (2005).

\bibitem{BNP} The idea that the new-physics strong phases are
negligible was first discussed in A.~Datta and D.~London, Phys.\
Lett.\ B {\bf 595}, 453 (2004).

\bibitem{BKpi} For example, see S.~Baek, P.~Hamel, D.~London, A.~Datta
and D.~A.~Suprun, Phys.\ Rev.\ D {\bf 71}, 057502 (2005), and
references therein. 

\end{thebibliography}
\end{document}